\documentclass{article}

\usepackage{arxiv}

\usepackage[utf8]{inputenc} 
\usepackage[T1]{fontenc}    
\usepackage{hyperref}       
\usepackage{url}            
\usepackage{booktabs}       
\usepackage{amsfonts}       
\usepackage{nicefrac}       
\usepackage{microtype}      
\usepackage{lipsum}		
\usepackage{graphicx}
\usepackage{natbib}
\usepackage{amsmath}
\usepackage{amsfonts}
\usepackage{doi}

\title{Learning from Imperfect Labels: A Physics-Aware Neural Operator with Application to DAS Data Denoising}

\author{ \href{https://orcid.org/0009-0008-4685-9458}{\includegraphics[scale=0.06]{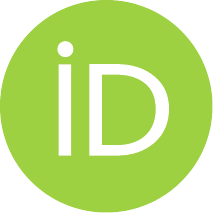}\hspace{1mm}Yang Cui}\\
	Department of Geosciences\\
	King Fahd University of Petroleum and Minerals\\
	Dhrhran, Saudi Arabia \\
	\texttt{yang.cui512@gmail.com} \\
	\And
	\href{https://orcid.org/0000-0002-4729-2659}{\includegraphics[scale=0.06]{orcid.pdf}\hspace{1mm}Denis Anikiev} \\
	Center for Integrative Petroleum Research (CIPR)\\
	King Fahd University of Petroleum and Minerals\\
	Dhrhran, Saudi Arabia \\
	\texttt{denis.anikiev@kfupm.edu.sa} \\
	\And
	\href{https://orcid.org/0000-0002-5189-0694}{\includegraphics[scale=0.06]{orcid.pdf}\hspace{1mm}Umair Bin Waheed} \\
	Department of Geosciences\\
	King Fahd University of Petroleum and Minerals\\
	Dhrhran, Saudi Arabia \\
	\texttt{umair.waheed@kfupm.edu.sa} \\
	\And
	\href{https://orcid.org/0000-0001-6429-4261}{\includegraphics[scale=0.06]{orcid.pdf}\hspace{1mm}Yangkang Chen} \\
	Bureau of Economic Geology\\
	The University of Texas at Austin\\
    Austin, TX, USA\\
	\texttt{chenyk2016@gmail.com} \\
}

\renewcommand{\shorttitle}{Physics-Aware DAS Denoising}

\hypersetup{
pdftitle={SeisDNO},
pdfsubject={q-bio.NC, q-bio.QM},
pdfauthor={David S.~Hippocampus, Elias D.~Striatum},
pdfkeywords={First keyword, Second keyword, More},
}

\begin{document}
\maketitle

\begin{abstract}
Supervised deep learning methods typically require large datasets and high-quality labels to achieve reliable predictions. However, their performance often degrades when trained on imperfect labels. To address this challenge, we propose a physics-aware loss function that serves as a penalty term to mitigate label imperfections during training. In addition, we introduce a modified U-Net–Enhanced Fourier Neural Operator (UFNO) that achieves high-fidelity feature representation while leveraging the advantages of operator learning in function space. By combining these two components, we develop a physics-aware UFNO (PAUFNO) framework that effectively learns from imperfect labels. To evaluate the proposed framework, we apply it to the denoising of distributed acoustic sensing (DAS) data from the Utah FORGE site. The label data were generated using an integrated filtering-based method, but still contain residual coupling noise in the near-wellbore channels. The denoising workflow incorporates a patching-based data augmentation strategy, including an uplifting step, spatial-domain convolutional operations, spectral convolution, and a projection layer to restore data to the desired shape. Extensive numerical experiments demonstrate that the proposed framework achieves superior denoising performance, effectively enhancing DAS records and recovering hidden signals with high accuracy. Codes and data related to this work are fully open-sourced via~\url{https://github.com/cuiyang512/PAUFNO} (The repository is private now; we will make it public upon acceptance of this paper).

\end{abstract}

\keywords{Physics-Aware Deep Learning \and Neural Operator \and DAS \and Denoising}

\section{Introduction}
Distributed Acoustic Sensing (DAS) is an emerging technique that enables the acquisition of ultra-high-density seismic records, offering a cost-effective and high-resolution approach to seismic exploration. In seismology, DAS has been successfully deployed in various applications, such as well logging, reservoir monitoring, $CO_2$ storage monitoring, and earthquake monitoring, positioning it as a viable alternative to conventional geophone-based approaches~\citep{FichtnerWalterEtAl2025}. Compared with traditional seismic acquisition methods, DAS offers several unique advantages, including resistance to high temperatures and pressures, ultra–high channel density, and exceptional sensitivity. These features allow DAS to deliver high-resolution seismic data even under extreme environmental conditions.

However, this very sensitivity also makes DAS particularly susceptible to noise, which complicates subsequent data processing and interpretation. The noise manifests in diverse and complex forms, such as random noise, erratic noise, horizontal noise, chessboard noise, fiber system noise, and zigzag noise, posing significant challenges to accurate seismic imaging. Consequently, developing effective denoising strategies for DAS data remains a critical and ongoing research priority. Over the past few decades, numerous denoising methods have been proposed for seismic data. These approaches can be broadly categorized into mathematical model–driven techniques and deep learning (DL)–based, data-driven methods~\citep{mousavi2022deep}. Despite their effectiveness, most existing techniques are designed to suppress only a single type of noise (e.g., random noise, erratic noise, or surface waves) within seismic profiles, limiting their applicability in real-world DAS scenarios. The ultra-dense nature of DAS records provides significant advantages for DL-based processing, as deep learning excels at performing complex, nonlinear, high-order mappings across massive datasets. In contrast, conventional filter-based methods are computationally inefficient and tend to perform poorly when handling DAS records with strong coherent noise~\citep{cui2025unsupervised}.

Over the past decades, DL has drawn significant attention in the seismology community~\citep{mousavi2022deep}, with models like Convolutional Neural Networks (CNNs), Autoencoders (AEs), Long Short-Term Memory (LSTM), U-Net, Transformers, and Kolmogorov–Arnold Networks (KANs) typically serving as discriminative models, while Generative Adversarial Networks (GANs) and Diffusion Models represent the class of generative models. Discriminative models have demonstrated high accuracy in seismic phase detection, reconstruction, and inversion by effectively capturing complex nonlinear relationships between the training and target domains~\citep{han2024mamcl, cui2025earthquake}. However, they often show poor robustness when even training with large-scale domain-specific data. In contrast, generative models excel at synthesizing realistic seismic records, facilitating denoising, super-resolution, and inversion, though their training is often computationally demanding and may lack interpretability~\citep{dou2023mda, durall2023deep}. Bearing the principles of discriminative and generative models in mind, we introduce Fourier Neural Operators (FNOs)~\citep{li2020fourier, wen2022u}, which have emerged as a promising paradigm in seismology. Unlike conventional discriminative or generative approaches, FNOs learn mappings directly between function spaces, enabling efficient modeling of wavefield propagation and other PDE-governed processes~\citep{zhu2023fourier, song2024seismic, badawi2025neural, ma2025effective}. In the context of seismic denoising, this unique property allows FNOs to capture global frequency-domain features for effective suppression of diverse and complex noise types, while simultaneously preserving the physical consistency of seismic wavefields. \cite{xie2025simultaneous} proposed a physics-informed neural network to suppress terratic and random noise, which is highlighted by embedding physics information for seismic denoising. As a result, FNOs provide superior generalization to unseen scenarios and offer a robust, computationally efficient, and physically grounded framework for seismic data processing, particularly for challenging DAS denoising tasks.

\subsection{Related work}
In practice, processing techniques developed for geophone-based seismic data are often applied to DAS acquisitions. Nevertheless, the unique characteristics of DAS pose challenges for model-specific denoising. For instance, \cite{Lellouch2021LowMagnitude} employed both low-pass and median filters to suppress strong noise in DAS records. Additionally, trace editing is commonly applied to handle severely noisy traces. To mitigate these issues, \cite{chen2023denoising} proposed the integrated denoising framework (IDF), a method that integrates multiple filters to suppress various types of noise. Specifically, the bandpass filter removes high-frequency noise, the structure-oriented median filter further suppresses high-amplitude erratic noise, and the dip filter in the $f-k$ domain attenuates regular vertical and horizontal noise. However, the high computational cost of IDF limits its applicability to large-scale datasets. \cite{oboue2024protecting} enhanced the structure-oriented median filter through a spatially varying scheme and introduced a local orthogonalization (LO) approach to suppress strong noise while preserving weak signals, which shows robust denoising performance across different tests from the Utah FORGE site~\citep{GDR_Dataset_1185}. 

DL is characterized by learning a non-linear mapping from massive data, which makes it well-suited for processing DAS records. Recently, numerous DAS denoisers based on both supervised learning (SL) and unsupervised learning (USL) methods have been developed. \cite{van2021self} proposed a self-supervised learning method for denoising DAS data with complex noise, while \cite{dong2022multiscale} introduced a multiscale spatial attention network to eliminate background noise, enhancing the detection of subtle signals by prioritizing local feature extraction in DAS data. Similarly, \cite{wu2022multi} developed a multiscale denoising framework utilizing convolutional operations across various data dimensions. \cite{zhong2023multi} presented a multiscale encoder-decoder network aimed at both background noise suppression and weak signal recovery. \cite{yang2023denoising} proposed a fully convolutional framework to extract hidden signals and attenuate diverse noise types in DAS data. \cite{yang2023SLKNet} also introduced SLKNet, an SL method for DAS-VSP noise suppression, leveraging a dense-connection network with a kernel-wise attention mechanism. \cite{zhu2023diffusion} developed a diffusion model to mitigate strongly coupled noise in DAS-VSP data, generating training data using the wave equation to simulate noise-free labels and an existing algorithm~\citep{yu2018borehole} to produce coupling noise. Building on the work of \cite{zhu2023diffusion}, \cite{xu2024vsp} trained a U-net model to suppress strong ``zigzag'' noise, although they relied exclusively on synthetic data for training. \cite{birnie2024explainable} employed a blind-mask self-supervised network to denoise various types of coherent noise in seismic profiles. Despite the notable success of these approaches in both geophone and DAS data denoising, developing a robust, generalizable unsupervised method capable of handling diverse and unseen data remains an ongoing challenge. Based on our investigations, there are very few publications that utilize operator learning for seismic data denoising tasks, especially for DAS. Therefore, we aim to explore its potential for suppressing diverse types of noise in DAS data as a case study. 

\begin{figure}
    \centering
    \includegraphics[width=1\linewidth]{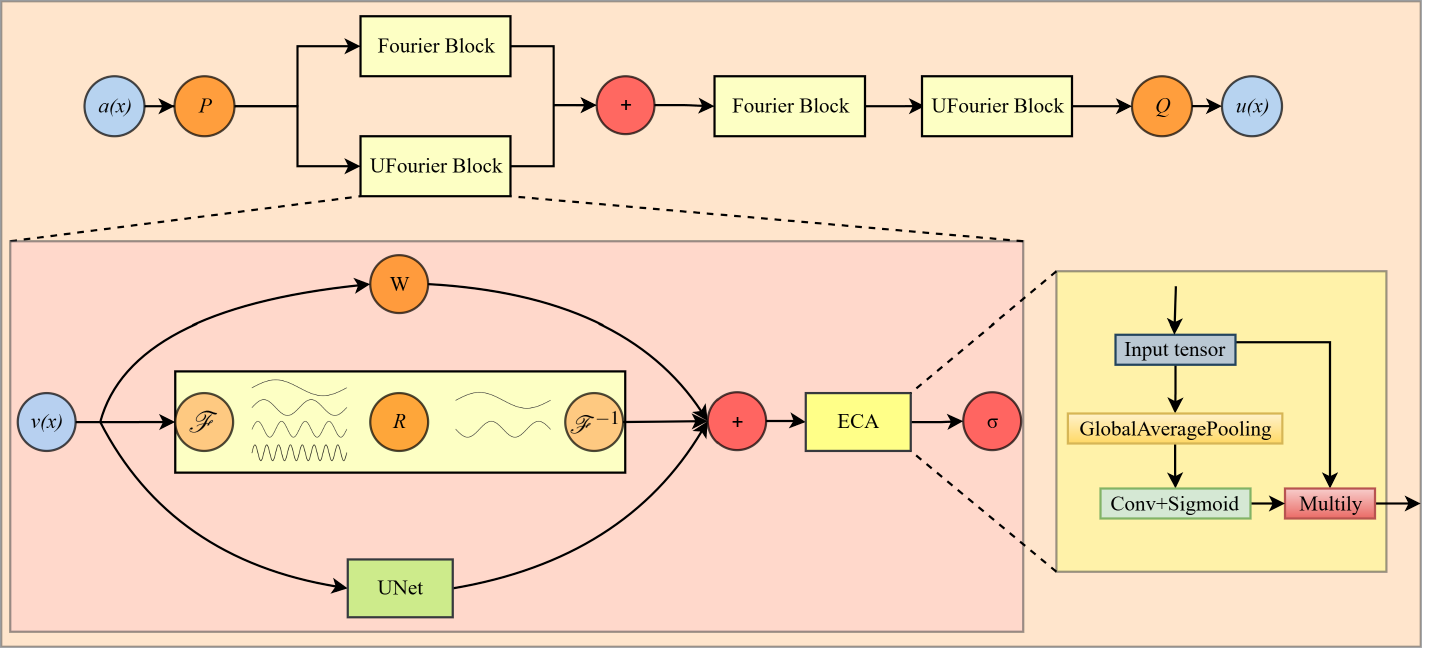}
    \caption{Diagram of the proposed DAS denoising framework. The model begins with an uplifting layer ($P$) that projects the input features into a higher-dimensional space. The resulting feature maps are then processed in parallel by the UFNO and FNO blocks, followed by an efficient channel attention (ECA) module that captures fine-grained information in both the spatial and frequency domains. Unlike the standard FNO block, the UFNO branch includes an additional U-Net channel to enhance feature representation in the time domain. Finally, a projection layer ($Q$) maps the high-dimensional features back to the input dimensionality to produce the denoised output. The symbol $\sigma$ denotes the activation function.}
    \label{fig:network}
\end{figure}

\begin{figure}
    \centering
    \includegraphics[width=1\linewidth]{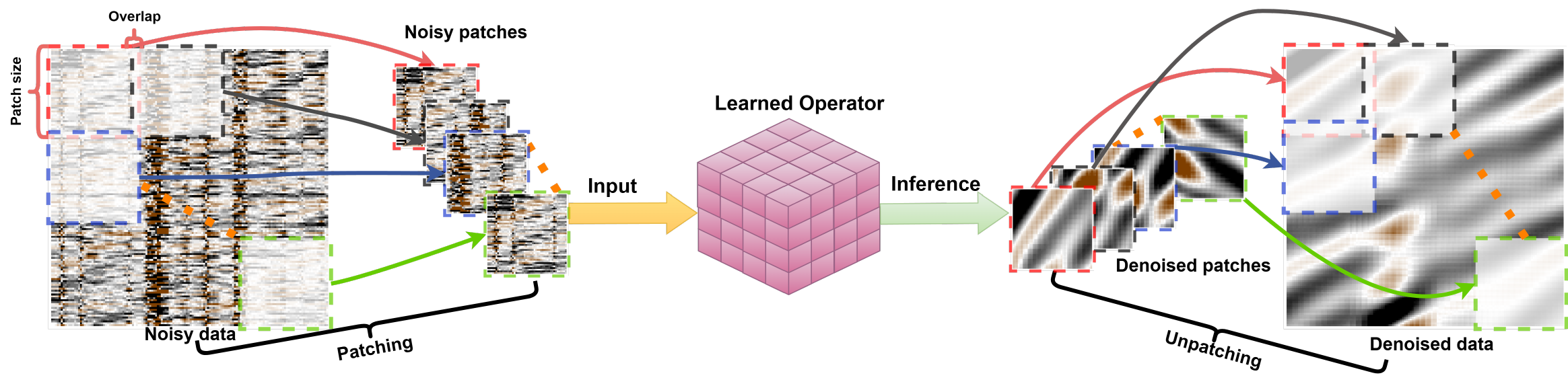}
    \caption{Inference workflow of the trained operator. The input noisy records are first divided into overlapping patches with fixed dimensions. These noisy patches are then processed by the trained operator to perform denoising. After obtaining the denoised patches, an inverse patching scheme is applied to reconstruct the full denoised output. It is worth noting that the trained operator can effectively handle patches with resolutions different from those used during training, as it learns underlying functional relationships rather than fixed nonlinear mappings within finite domains.}
    \label{fig:inference}
\end{figure}

Many researchers have investigated DAS denoising using open-source data from the Utah FORGE site~\citep{yang2023denoising,yang2023SLKNet,saad2024unsupervised,xu2024selfmixed,cui2025unsupervised}. However, none of these studies have addressed the presence of coupling noise in near-wellbore channels. The existence of coupling noise in the near-wellbore channels (see Figure~\ref{fig:eq_74_seis}) can adversely affect subsequent data-processing steps, such as arrival-time picking and microseismic event localization. Therefore, developing an approach capable of further suppressing the residual noise in the label data is of critical importance. To address this issue, we propose a U-Net–enhanced Fourier Neural Operator (UFNO) that integrates a physics-aware loss function to effectively attenuate strong noise in DAS recordings (see Figure~\ref{fig:network}). PAUFNO combines the advantages of operator learning and physics-informed neural networks, enabling it to significantly outperform conventional model-driven approaches and the U-Net baseline. Unlike CNNs or U-Nets that learn finite-domain mappings, the FNO learns an operator between function spaces, allowing inference at resolutions different from those used during training. We employ a patching strategy~\citep{chen2020fast} to extract 2D training patch pairs from the Utah FORGE field dataset.  Although the model is trained on imperfect labels containing residual coupling noise, the incorporation of the physics-aware loss function mitigates the influence of these imperfections. Once trained, the learned operator can generalize to various input types and resolutions using the workflow illustrated in Figure~\ref{fig:inference}, without the need for retraining. Numerical experiments further confirm the effectiveness and robustness of the proposed denoising framework.

The remainder of this paper is organized as follows. We first introduce the principles of the proposed PAUFNO framework and its corresponding DAS denoising workflow. Next, we present a comparative evaluation of denoising performance between PAUFNO and several benchmark approaches on field datasets collected from different surveys. We then discuss the model’s generalization capability, uncertainty quantification, and limitations. Finally, we summarize the proposed DAS denoising strategy, which effectively mitigates coupling noise arising from imperfect fiber–well coupling by incorporating physics-aware loss function.

\section{Methodology}\label{method}
This section first introduces the problem formulation for DAS denoising, followed by the diagram and operational principles of the proposed neural operator (NO)–based denoising framework. Finally, we present the physics-aware loss function designed to enhance coupling-noise suppression.

\subsection{Problem setup}
During the data acquisition process, sensors such as geophones and fiber-optic cables inevitably capture both signal and noise. Furthermore, due to the high sensitivity of DAS cables, they are particularly prone to recording additional noise sources, including fiber system noise, coupling noise arising from poor contact between the DAS fiber and the wellbore wall or ground surface, and ambient noise. The acquired DAS records can be mathematically formulated as:
\begin{equation}
    \label{eq:denoising}
    \mathbf{Y}=\mathbf{S}+\mathbf{N}+\mathbf{r},
\end{equation}
where $\mathbf{Y}$ denotes the recorded DAS data, $\mathbf{S}$ the desired signal component, $\mathbf{N}$ the noise component, and $\mathbf{r}$ the residual noise. The goal of denoising is to recover the underlying signal $\mathbf{S}$ from the field data $\mathbf{Y}$ while effectively suppressing the variable noise $\mathbf{N}$. In supervised deep learning, large amounts of labeled data are required for training. However, obtaining accurate ground-truth labels for field data is inherently challenging, which often leads to residual noise $\mathbf{r}$ that remains difficult to eliminate completely. The spectral energy distribution differences of signal and coupling noise provide the basis for designing the FK masks in the next sections.

\subsection{Network principle}
The proposed network builds on the Fourier Neural Operator (FNO)~\citep{li2020fourier}, which learns nonlinear mappings between infinite-dimensional function spaces and has proven effective for solving partial differential equations (PDEs). In this work, we repurpose this operator-learning paradigm for DAS denoising, a non-PDE setting, to explore the broader applicability of neural operators (NOs).

As shown in Figure~\ref{fig:network}, the architecture comprises an input uplifting layer $P$, a stack of FNO and U-FNO blocks augmented with Efficient Channel Attention (ECA)~\citep{Wang_2020_CVPR_ECA}, and an output projection layer $Q$. The lifting layer $P$ expands the feature dimensionality from 1 channel to $\texttt{width}$, enhancing representational capacity. The subsequent operator blocks extract complementary information by combining global spectral modeling (via truncated Fourier transforms) with local convolutional pathways; in U-FNO blocks, a U\,-Net branch further refines locality while ECA adaptively reweights channels to emphasize informative responses. Finally, the projection layer $Q$ maps the features from the lifted space back to the input dimensionality, yielding a denoised estimate with the same spatial resolution as the input.

\subsection{Physics-aware loss function}
In the context of DAS denoising, where fiber-optic coupling artifacts often manifest as high-energy noise in specific low-velocity cones within the frequency-wavenumber (FK) domain. The residual high-energy coupled noise is hard to suppress, as indicated by~\cite{chen2023denoising}. Furthermore, prior research trained on label data generated by the integrated filters proposes that a hybrid supervised loss function is employed to balance data fidelity with physics-aware constraints. 

\begin{figure}
    \centering
    \includegraphics[width=1\linewidth]{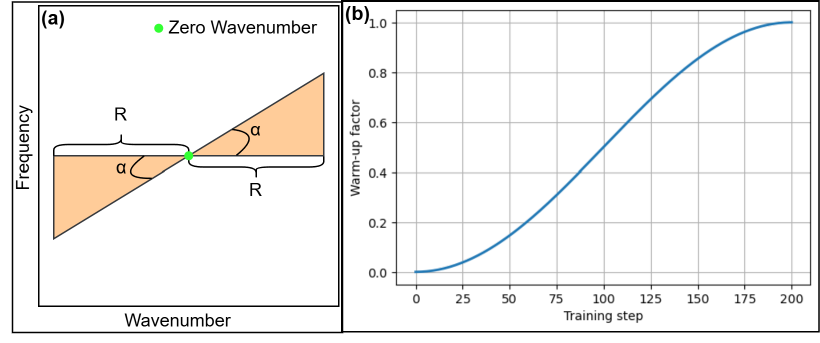}
    \caption{Diagram of the physics-aware loss function. (a) Dual-triangle masking setup, where the green dot indicates the zero wavenumber point, as well as the zero point of the normalized frequency. (b) Warmup factor of the physics-aware loss, where the factor rises from 0 to 1 to avoid useful signal damages in the early training stage.}
    \label{fig:warmup_factor}
\end{figure}

This approach leverages the dual-triangular penalty mask (shown in Figure~\ref{fig:warmup_factor}a) to suppress noise in targeted FK regions while preserving seismic signal integrity. Specifically, let $\mathbf{x} \in \mathbb{R}^{B \times C \times H \times W}$ denote the predicted denoised time-distance data, $\mathbf{y} \in \mathbb{R}^{B \times C \times H \times W}$ the target labels with residual coupling noise, $\mathbf{M} \in \mathbb{R}^{H \times W}$ the precomputed triangular penalty mask. Figure~\ref{fig:warmup_factor}a exhibits the proposed dual-triangle masks, where we keep the values inside as 1 and outside as 0. The size of the dual-triangle mask is designed by the energy distribution of coupling noise in FK spectra. By analyzing the residual coupled noise and the signal’s FK spectrum, we ultimately selected a dual-triangular mask. As illustrated in Figure~\ref{fig:warmup_factor}a, the vertices of the two triangles are positioned at the zero-wavenumber point. The short-side vertex angle $\alpha$ is set to $35^\circ$, and the leg adjacent to this $35^\circ$ angle has a length of 40 sample points. The shape of the triangular mask is designed according to the energy distribution of the noise. Our goal is to cover the noise energy region as fully as possible while minimizing contamination of the signal energy. Moreover, we discussed the selection of the short-side vertex angle $\alpha$ in the ``DISCUSSION'' section.

As shown in Figure~\ref{fig:warmup_factor}b, $s$ denotes the current training step, $s_w$ the number of warmup steps (set to 200 by default, consistent with the number of epochs), and $1-\lambda$ represents the hyperparameter that balances the FK term. In the hybrid loss function (Equation~\ref{eq:total_loss}), we set $\lambda=0.98$. This warmup scheme for the FK loss introduces a soft energy cutoff, thereby preventing signal degradation in the initial training phase.

The FK dual-triangle loss, $\mathcal{L}_\text{FK}(\mathbf{x}, s)$, penalizes excessive energy within these noise-prone triangular regions of the FK spectrum, formulated as
\begin{equation}
\label{eq:fk_loss}
\mathcal{L}_{FK} (x, s) = w(s) \cdot \frac{1}{BCHW} \sum_{b=1}^B \sum_{c=1}^C \sum_{h=1}^H \sum_{w=1}^W \left[ M_{h,w} \cdot \left| \mathcal{F}(x_{b,c}) \right|_{h,w} \right]^2,
\end{equation}
where $\mathcal{F}(\cdot)$ represents the centered 2D fast Fourier transform defined by Equation~\ref{eq:fft} (unnormalized, with zero frequency centered), $|\cdot|$ is the complex modulus. Similar to the implementation in DASCore~\citep{chambers2024dascore}, we computed the FK spectra as follows:
\begin{equation}
\label{eq:fft}
\mathcal{F}(x_{b,c}) = \mathrm{fftshift}(\mathrm{fft2}(x_{b,c})),
\end{equation}

The cosine annealing warmup schedule (shown in Figure~\ref{fig:warmup_factor}b) is implemented that gradually ramps the penalty from 0 to 0.5 over $s_w$ steps, thereby avoiding premature over-constraint (causing signal damage) during the early training stage:
\begin{equation}
\label{eq:warm_up}
w(s) = 0.5 \left(1 - \cos\left(\pi \cdot \min\left(1, \frac{s}{s_w}\right)\right)\right),
\end{equation}

The FK loss proposed by Equations~\ref{eq:fk_loss} to~\ref{eq:warm_up} is to mitigate the remaining noise in labels caused by the poor coupling between the fiber optics and wellbore. Therefore, it acts as a correction term in the training loop. However, to maintain the correct parts in labels and enforce the pixel-wise reconstruction accuracy, we need to employ the supervised loss term as shown in Equation~\ref{eq:l1loss}:
\begin{equation}
\label{eq:l1loss}
\mathcal{L}_{L1} (x, y) = \frac{1}{BCHW} \sum_{b=1}^B \sum_{c=1}^C \sum_{h=1}^H \sum_{w=1}^W \left[ x_{b, c, h,w}-y_{b, c, h,w} \right],
\end{equation}

Therefore, we get the total hybrid loss, $\mathcal{L}_\text{total}(\mathbf{x}, \mathbf{y}, s)$, integrates these components via a weighted sum:
\begin{equation}
\label{eq:total_loss}
\mathcal{L} = \lambda \cdot \mathcal{L}_{L1} (x, y) + (1 - \lambda) \cdot \mathcal{L}_{FK} (x, s)
\end{equation}
which simultaneously minimizes reconstruction discrepancies and attenuates noise energy in the FK cones, enhancing DAS data quality by mitigating coupling-induced artifacts without distorting valid wavefields.

\subsection{Uncertainty quantification}
MC Dropout has been widely recognized as an effective approach for quantifying uncertainty in deep learning models, particularly when assessing the confidence of model predictions. In this study, we employ the MC Dropout technique to evaluate the denoising confidence of the PAUFNO model, with a specific focus on the near-wellbore region (channels 0–50), where the loss term enforces correction. 

The model incorporates dropout layers within the convolutional and U-Net components, which are reactivated during inference to generate a set of stochastic predictions. Specifically, for each input sample $\mathbf{x}$, the model produces $T$ stochastic predictions 
$\{ \hat{\mathbf{y}}_t \}_{t=1}^{T}$ by applying dropout masks during inference. The predictive mean and uncertainty are then computed as:
\begin{equation}
\bar{\mathbf{y}} = \frac{1}{T} \sum_{t=1}^{T} \hat{\mathbf{y}}_t,
\label{eq:mc_dropout}
\end{equation}
where $\bar{\mathbf{y}}$ represents the ensemble mean, which serves as the denoised results.
\begin{equation}
\sigma = \sqrt{ \frac{1}{T} \sum_{t=1}^{T} (\hat{\mathbf{y}}_t - \bar{\mathbf{y}})^2 },
\label{eq:mc_dropout_1}
\end{equation}
where $\sigma$ quantifies the model's epistemic uncertainty. Higher Standard Deviation (STD) values indicate regions of low model confidence, often corresponding to areas with low signal-to-noise ratio and complex signal distribution.

\section{Results}
In this section, we introduce the operator inference results, comparing the results of trained operators with those of the IDF~\citep{chen2023denoising} method. Moreover, we will introduce the out-of-distribution training data to validate the generalizability of the trained physics-aware neural operator.

\subsection{Training data generation}
The fiber-optic cable was installed in a borehole configuration to serve as a dense receiver array in the monitoring well 78-32. The data were recorded with a sampling interval of 0.5~ms, a channel spacing of 1~m, and a gauge length of 10~m. All labels were generated using the integrated, filter-based denoising method proposed by~\cite{chen2023denoising}. After data augmentation, this process yielded 1,152 training pairs and 128 validation pairs, each of size $128\times128$. To ensure reproducibility, we provide all hyperparameters used in training in Table~\ref{tab:hyperparameters}. Moreover, to provide a comprehensive comparison, we trained the proposed UFNO in a data-driven manner together with baseline U-Net~\citep{ronneberger2015u} to benchmark PAUFNO. All hyperparameters, i.e., learning rate, patch size, number of training samples, and epochs, were kept identical across models to ensure a fair comparison. All experiments were conducted on a workstation equipped with an NVIDIA RTX A4500 GPU. In addition, we compare PAUFNO with IDF to demonstrate its effectiveness relative to an integrated filtering-based approach. Table~\ref{tab:training_efficience} summarizes the training time, number of trainable parameters, and inference time for a single example (with a spatial dimension of $2000 \times 960$ samples) from the Utah FORGE site. It is evident that PAUFNO and UFNO contain fewer trainable parameters than the U-Net. However, both models require longer training and inference times than the U-Net due to their feature representations in the Fourier domain are time-consuming.

\begin{table}[htbp]
    \centering
    \caption{Hyperparameter configuration}
    \begin{tabular}{ll}
        \toprule[1.5pt]
        \textbf{Hyperparameter} & \textbf{Specification} \\
        \midrule[1.5pt]
        Optimizer              & Adam                    \\
        Epoch                  & 200                     \\
        Learning Rate          & $5 \times 10^{-4}$      \\
        Loss Function          & L1 and FK Loss          \\
        Batch Size             & 128                     \\
        Patch Size             & $128 \times 128$        \\
        Modes                  & 12                       \\
        Width                  & 16                      \\
        Training Samples       & 1152                    \\
        Validation Samples     & 128                     \\
        \bottomrule[1.5pt]
    \end{tabular}
    \label{tab:hyperparameters}
\end{table}

\subsection{Utah FORGE examples}
\begin{figure}
    \centering
    \includegraphics[width=1\linewidth]{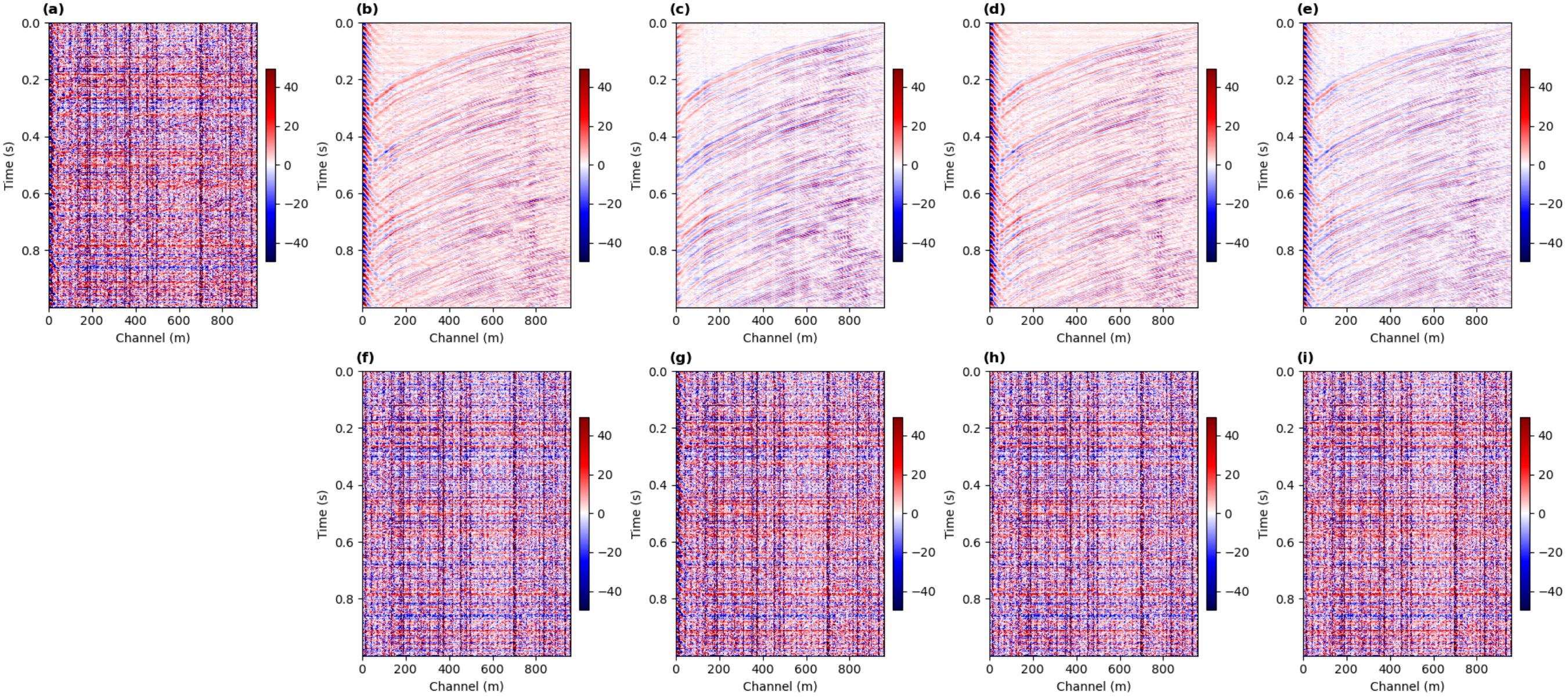}
    \caption{Denoising comparisons of different methods on DAS record with multiple events. (a) is the raw data. (b–e) are corresponding to denoised records using BPSOMFFK, PAUFNO, UFNO, and U-Net, respectively. (f–i) display removed noise with BPSOMFFK, Proposed, UFNO, and U-Net, respectively.}
    \label{fig:eq_36_seis}
\end{figure}

\begin{figure}
    \centering
    \includegraphics[width=1\linewidth]{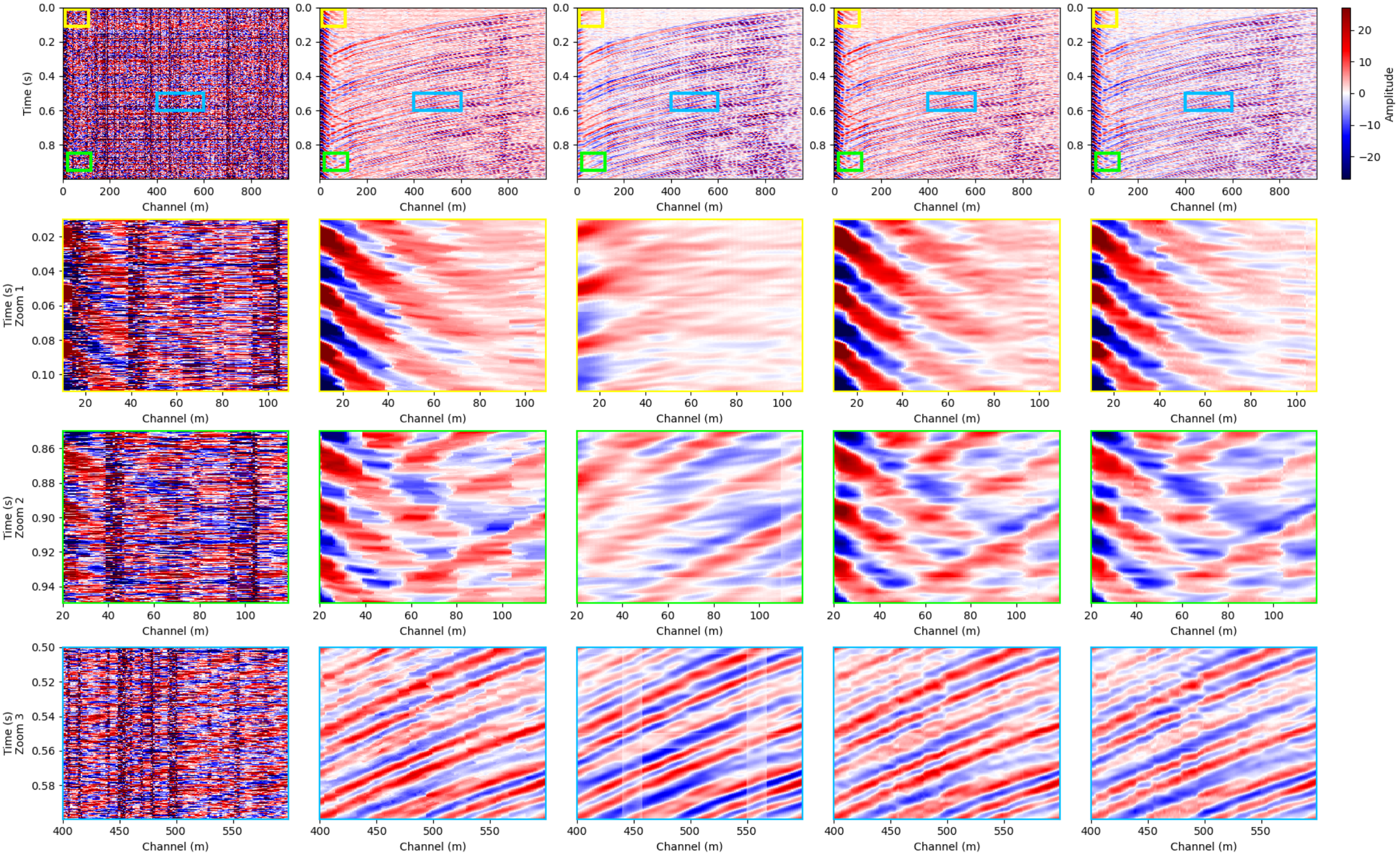}
    \caption{Zoomed-in comparisons of denoising results from Figure~\ref{fig:eq_36_seis}. The first row shows the raw data and denoising profiles using different methods. The yellow rectangles in the second row indicate regions affected by residual clipping noise, which are enlarged in the second row. The green rectangles highlight areas containing both coupling noise and signal, shown in the third row. The bottom row presents zoomed-in sections corresponding to regions dominated by signals only.}
    \label{fig:eq_36_zoomed}
\end{figure}

\begin{figure}
    \centering
    \includegraphics[width=1\linewidth]{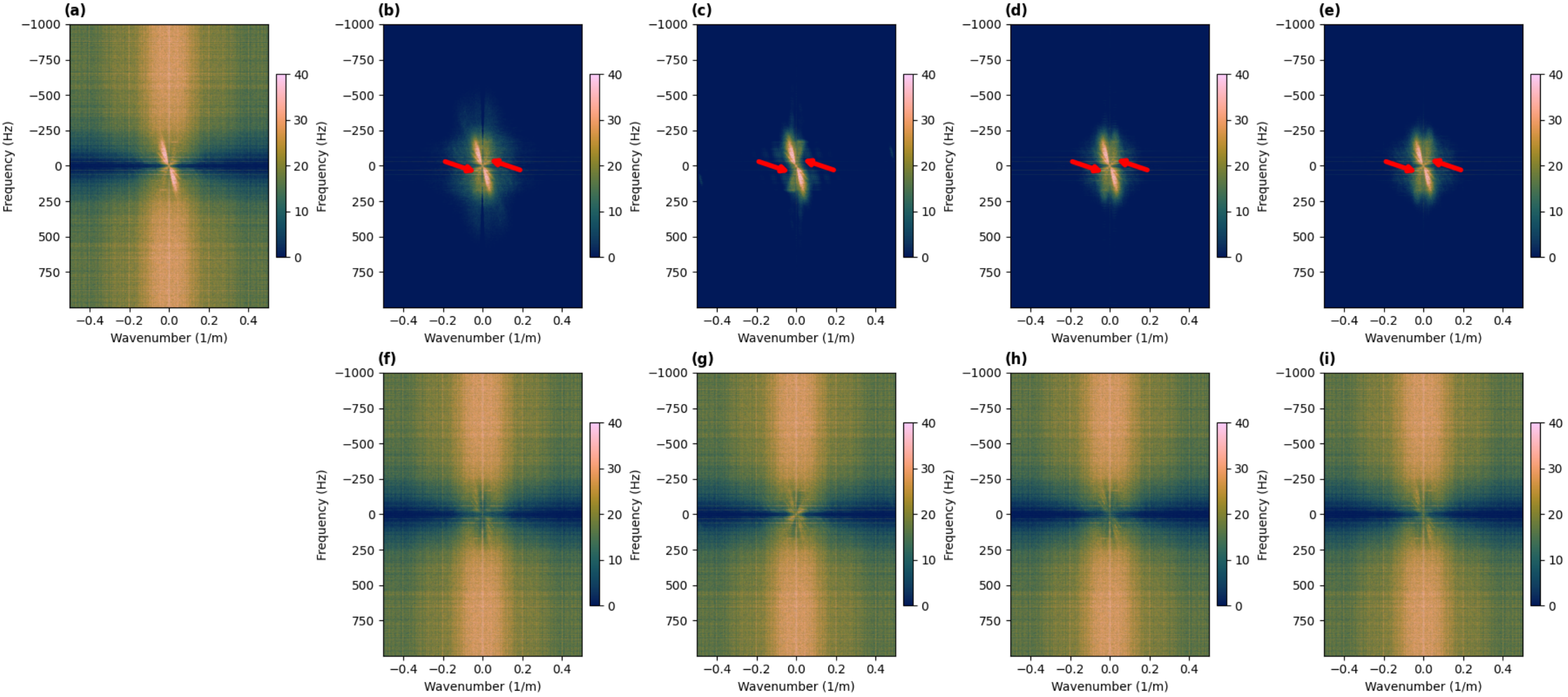}
    \caption{FK spectra comparisons of different methods on DAS record with multiple events. (a) is the spectra raw data. (b–e) are fk spectra corresponding to denoised records using BPSOMFFK, PAUFNO, UFNO, and U-Net, respectively. (f–i) display spectra of removed noise with BPSOMFFK, Proposed, UFNO, and U-Net, respectively. Note that we calculate the spectra using DASCore~\cite{chambers2024dascore}.}
    \label{fig:eq_36_spec}
\end{figure}

\begin{figure}
    \centering
    \includegraphics[width=1\linewidth]{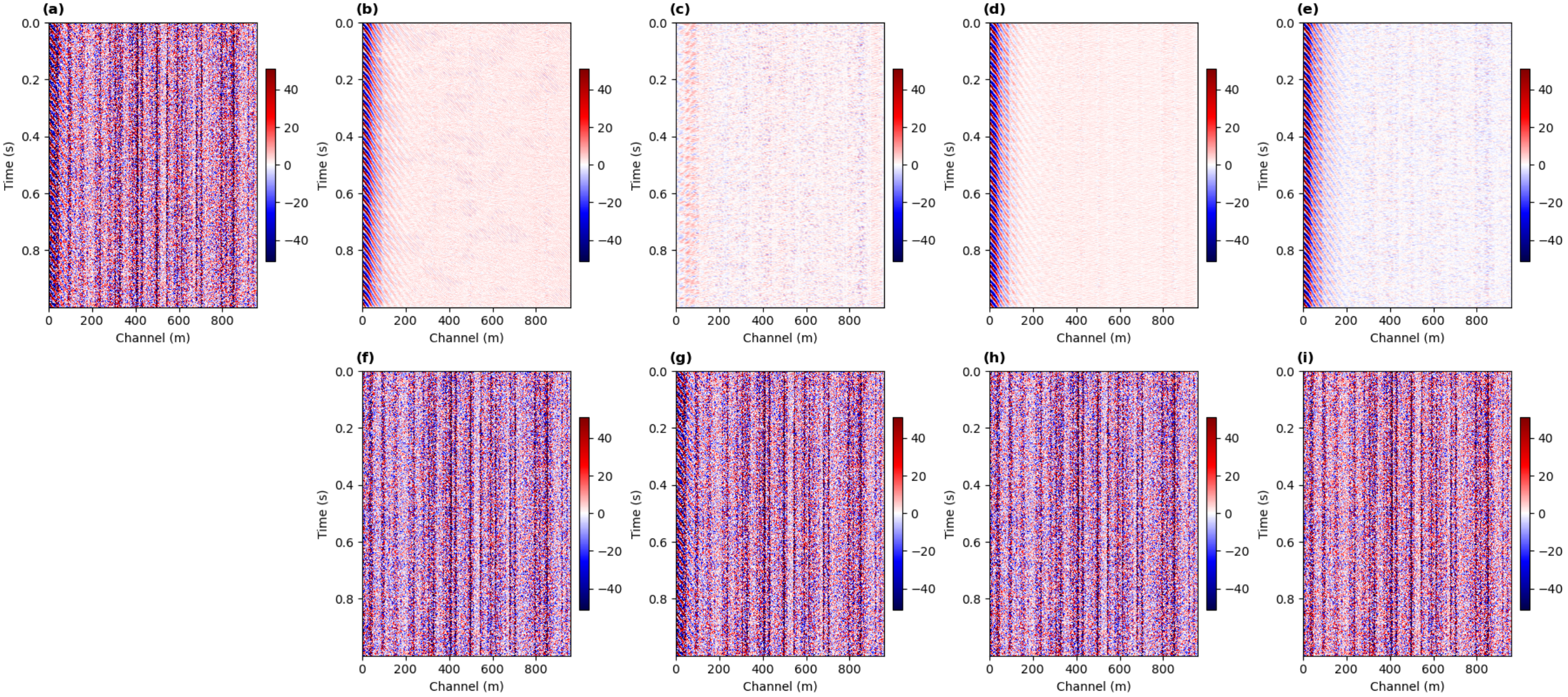}
    \caption{Denoising comparisons of different methods, and the DAS record is characterized by without any visible event. (a) is the raw data. (b–e) are corresponding to denoised records using BPSOMFFK, PAUFNO, UFNO, and U-Net, respectively. (f–i) display removed noise with BPSOMFFK, Proposed, UFNO, and U-Net, respectively.}
    \label{fig:eq_35_seis}
\end{figure}

\begin{figure}
    \centering
    \includegraphics[width=1\linewidth]{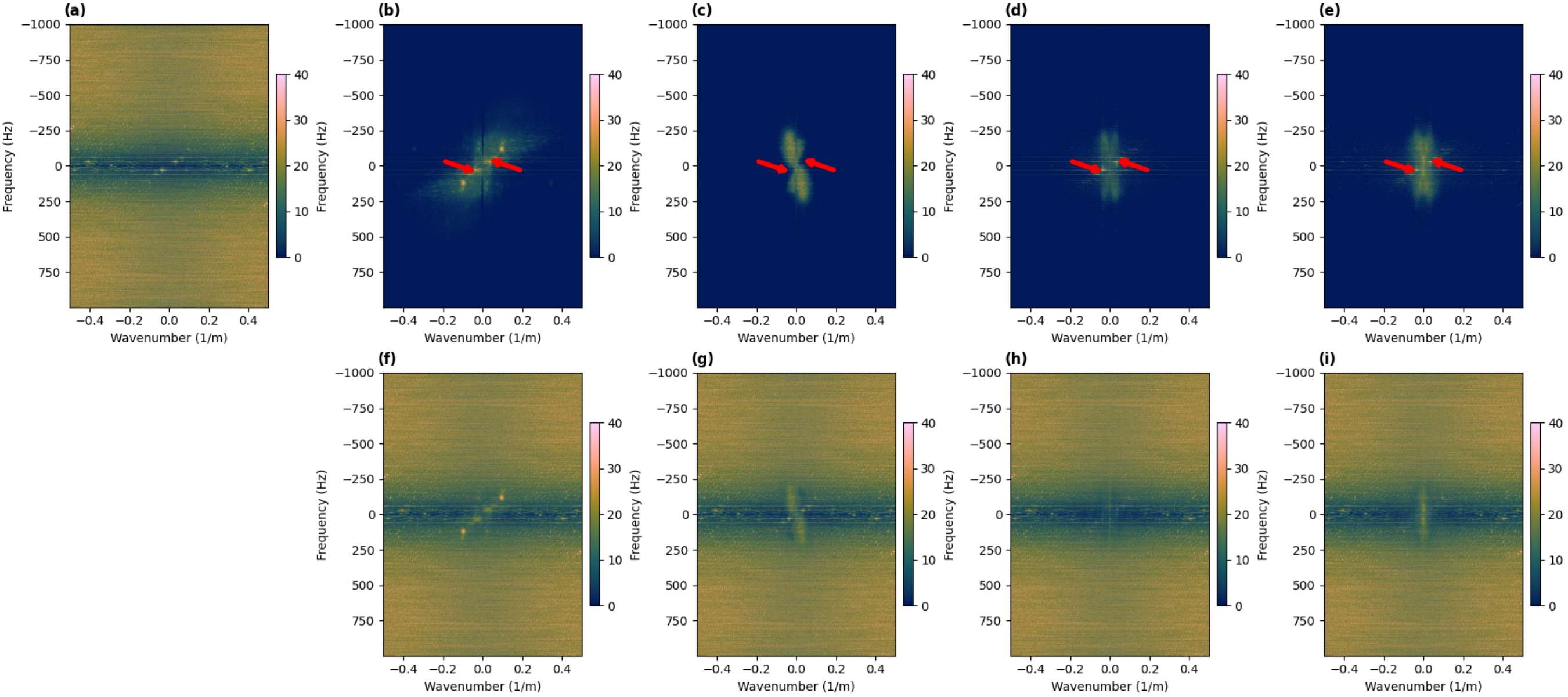}
    \caption{FK spectra comparisons of different methods on DAS record without events. (a) is the spectra raw data. (b–e) are fk spectra corresponding to denoised records using BPSOMFFK, PAUFNO, UFNO, and U-Net, respectively. (f–i) display spectra of removed noise with BPSOMFFK, Proposed, UFNO, and U-Net, respectively.}
    \label{fig:eq_35_spec}
\end{figure}

\begin{figure}
    \centering
    \includegraphics[width=1\linewidth]{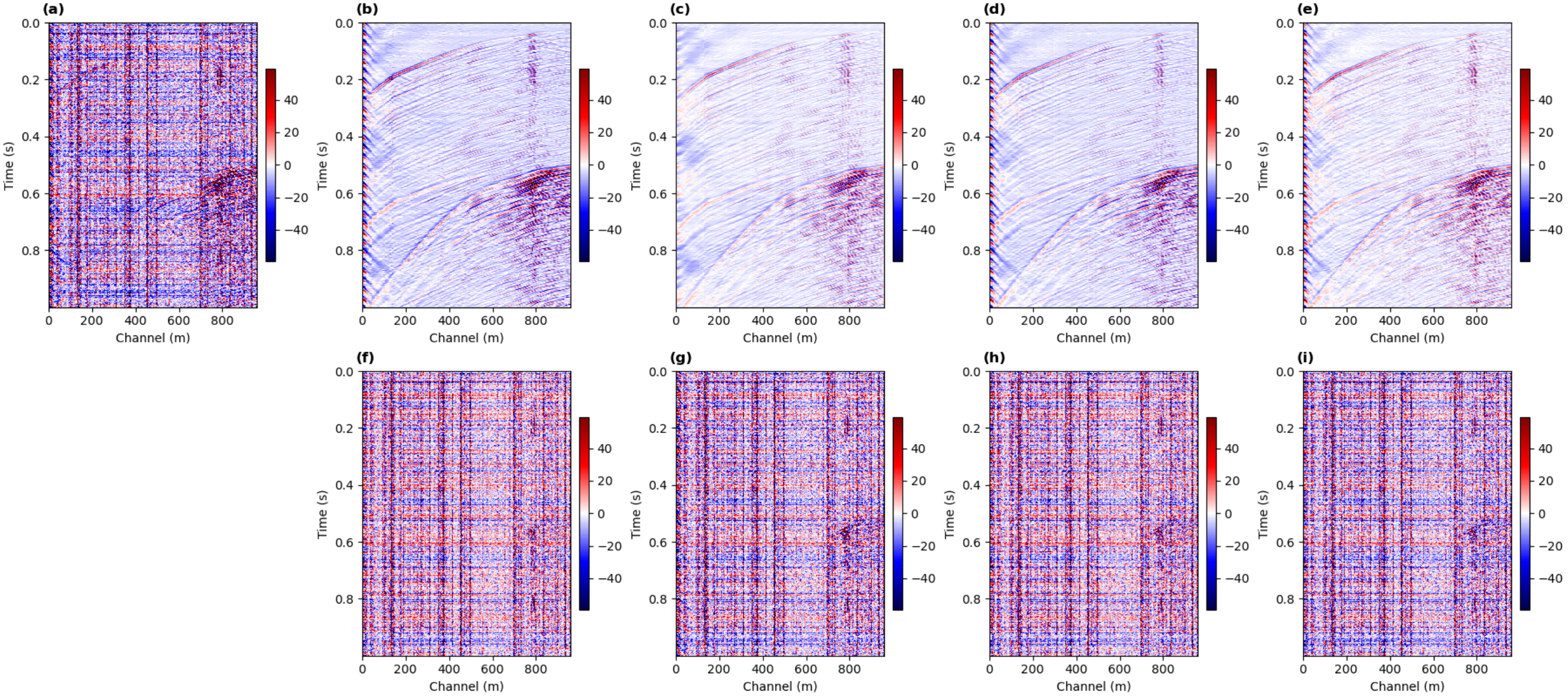}
    \caption{Denoising comparisons of different methods on DAS record with high-amplitude events. (a) is the raw data. (b–e) are corresponding to denoised records using BPSOMFFK, PAUFNO, UFNO, and U-Net, respectively. (f–i) display removed noise with BPSOMFFK, Proposed, UFNO, and U-Net, respectively.}
    \label{fig:eq_74_seis}
\end{figure}

\begin{figure}
    \centering
    \includegraphics[width=1\linewidth]{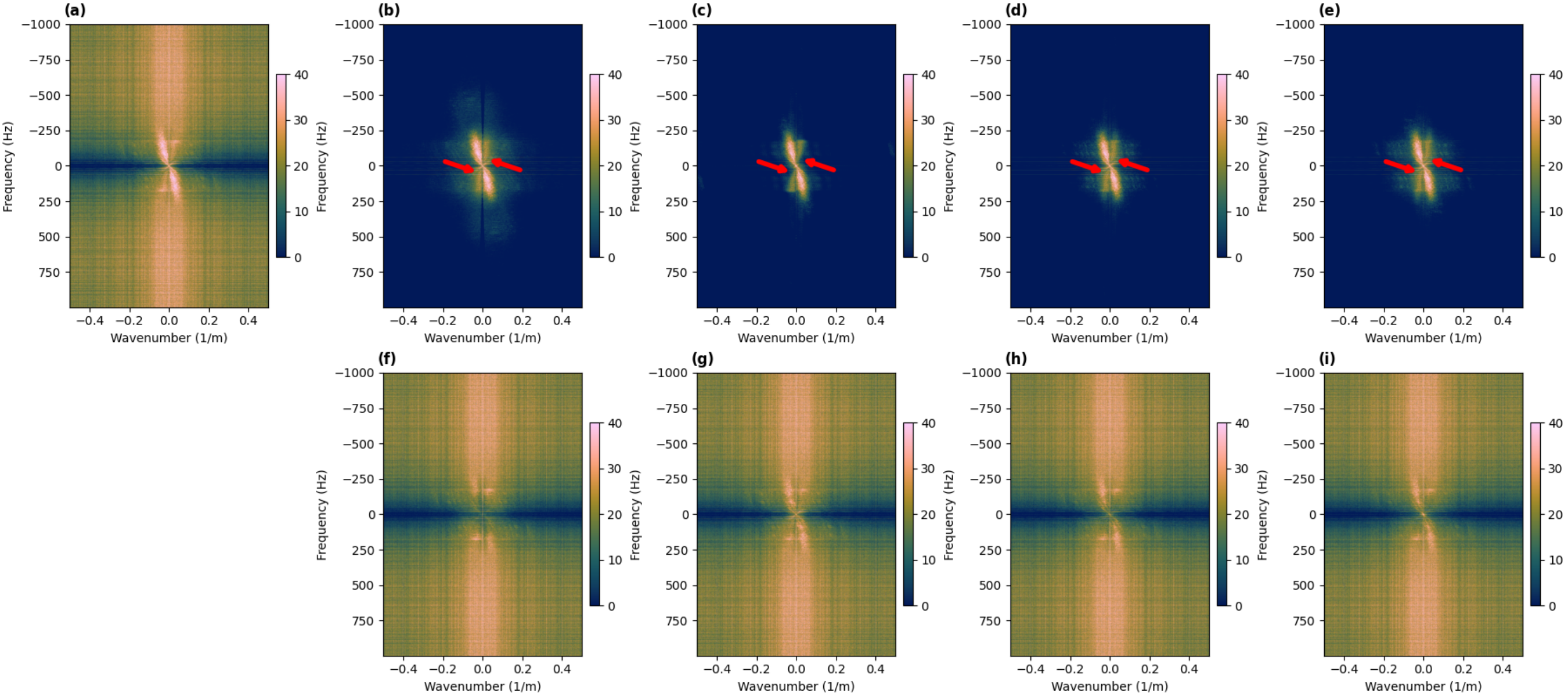}
    \caption{FK spectra comparisons of different methods on DAS record with high-amplitude events. (a) is the spectra raw data. (b–e) are fk spectra corresponding to denoised records using BPSOMFFK, PAUFNO, UFNO, and U-Net, respectively. (f–i) display spectra of removed noise with BPSOMFFK, PAUFNO, UFNO, and U-Net, respectively.}
    \label{fig:eq_74_spec}
\end{figure}

Figures~\ref{fig:eq_36_seis}–\ref{fig:eq_74_spec} present a comparative visualization of the denoising performance of PAUFNO against baseline approaches. The representative raw data shown in Figures~\ref{fig:eq_36_seis}a,~\ref{fig:eq_35_seis}a, and~\ref{fig:eq_74_seis}a exhibit distinct characteristics. For example, Figure~\ref{fig:eq_36_seis}a contains multiple seismic events, Figure~\ref{fig:eq_35_seis}a displays no visible events but strong residual coupling noise, and Figure~\ref{fig:eq_74_seis}a reveals hidden signals with high-amplitude energy. For better comparison, we also provide spectra comparisons in Figures~\ref{fig:eq_36_spec},~\ref{fig:eq_35_spec}, and~\ref{fig:eq_74_spec}. 

\begin{table}[htbp]
    \centering
    \caption{Comparison of training efficiency and inference performance among different methods.}
    \begin{tabular}{llll}
        \toprule[1.5pt]
        \textbf{Specification} & \textbf{Proposed} & \textbf{UFNO} & \textbf{U-Net}\\
        \midrule[1.5pt]
        Trainable Parameters    & 372,045   & 372,045   & 419,753  \\
        Training Time (s)           & 368.52    & 371.45    & 161.93   \\
        Inference Time (s)        & 3.91e-01    & 3.29e-01    & 4.28e-2   \\
        \bottomrule[1.5pt]
    \end{tabular}
    \label{tab:training_efficience}
\end{table}

As shown in Figure~\ref{fig:eq_36_seis}, all denoising approaches performed remarkably in terms of signal preservation and noise suppression. However, with the addition of the FK loss term, PAUFNO achieves superior performance in reducing residual noise caused by poor coupling between the fiber-optic cable and the wellbore. In particular, as shown in Figure~\ref{fig:eq_36_zoomed}, channels 0–50 demonstrate that incorporating the physics-aware loss function further suppresses coupling noise, even when the model is trained with imperfect labels. The zoomed-in sections in Figure~\ref{fig:eq_36_zoomed} correspond to regions containing only residual noise, both signal and coupled noise, and purely signal, respectively. These results clearly indicate that the physics-aware UFNO not only effectively recovers the underlying signals but also attenuates the coupled noise. To provide a more detailed comparison, the corresponding FK spectra of Figure~\ref{fig:eq_36_seis} are presented in Figure~\ref{fig:eq_36_spec}, which further highlights the effectiveness of the proposed dual-triangle penalty in the FK domain. As shown in Figures~\ref{fig:eq_36_spec}b and~\ref{fig:eq_36_spec}c, the proposed model markedly suppresses energy concentrated in the low-frequency and small-wavenumber regions (highlighted by red arrows), consistent with the removal of coupling noise observed in Figure~\ref{fig:eq_36_seis}c.

The second example (see Figure~\ref{fig:eq_35_seis}a), characterized by strong residual coupling noise and the absence of visible seismic events, is introduced to evaluate whether the denoising approaches generate artifacts when no hidden signal is present. The label data (Figure~\ref{fig:eq_35_seis}b) clearly exhibits persistent coupling noise, which leads to the failure of both the U-Net and data-driven UFNO in effectively removing it. In contrast, the PAUFNO significantly attenuates the residual coupling noise remaining in the label, demonstrating superior generalization and robustness under poor coupling conditions. As shown in Figures~\ref{fig:eq_35_spec}a–\ref{fig:eq_35_spec}e, the residual coupling noise exhibits spectral energy concentrated in the low-frequency and low-wavenumber regions, forming a trend from the upper right to the lower left. Moreover, the spectra shown in Figure~\ref{fig:eq_35_spec} further suggest that physics-aware UFNO achieves promising residual coupling noise suppression performance.

Figure~\ref{fig:eq_74_seis}a illustrates the DAS record of the poor coupling between the fiber-optical cable and wellbore condition, which requires specialized processing to mitigate. Compared with Figures~\ref{fig:eq_74_seis}b,~\ref{fig:eq_74_seis}d, and~\ref{fig:eq_74_seis}e, Figure~\ref{fig:eq_74_seis}c demonstrates that PAUFNO more effectively removes coupling noise while preserving useful signal components. The corresponding FK spectra (Figure~\ref{fig:eq_74_spec}) further confirm this result, showing that PAUFNO attenuates low-frequency and small-wavenumber energy with opposite dips to the hidden signals. In contrast to UFNO and U-Net, PAUFNO selectively reduces energy in the low-frequency and small-wavenumber regions (highlighted by red arrows) while retaining the spectral energy of the seismic signals.

\subsection{Generalization test}
We demonstrated the denoising performance of different methods on DAS recordings from the Utah FORGE site. Typically, for deep learning models trained in a supervised manner, it is necessary to retrain the network or apply transfer learning to achieve satisfactory performance on previously unseen data. To further assess their generalization and robustness, we conducted additional experiments on DAS data acquired from different field surveys. All models were trained exclusively on the Utah FORGE dataset and then applied to unseen data without any retraining. The unseen data were obtained from a 3D DAS-VSP survey conducted at the geothermal in-situ laboratory in Gro{\ss} Sch{\"o}nebeck, Germany~\citep{Martuganova2022_GSB_DASVSP}. The DAS measurements at this site were recorded with a temporal sampling interval of 2~ms and a spatial sampling interval of 5~m along the entire well.

\begin{figure}
    \centering
    \includegraphics[width=1\linewidth]{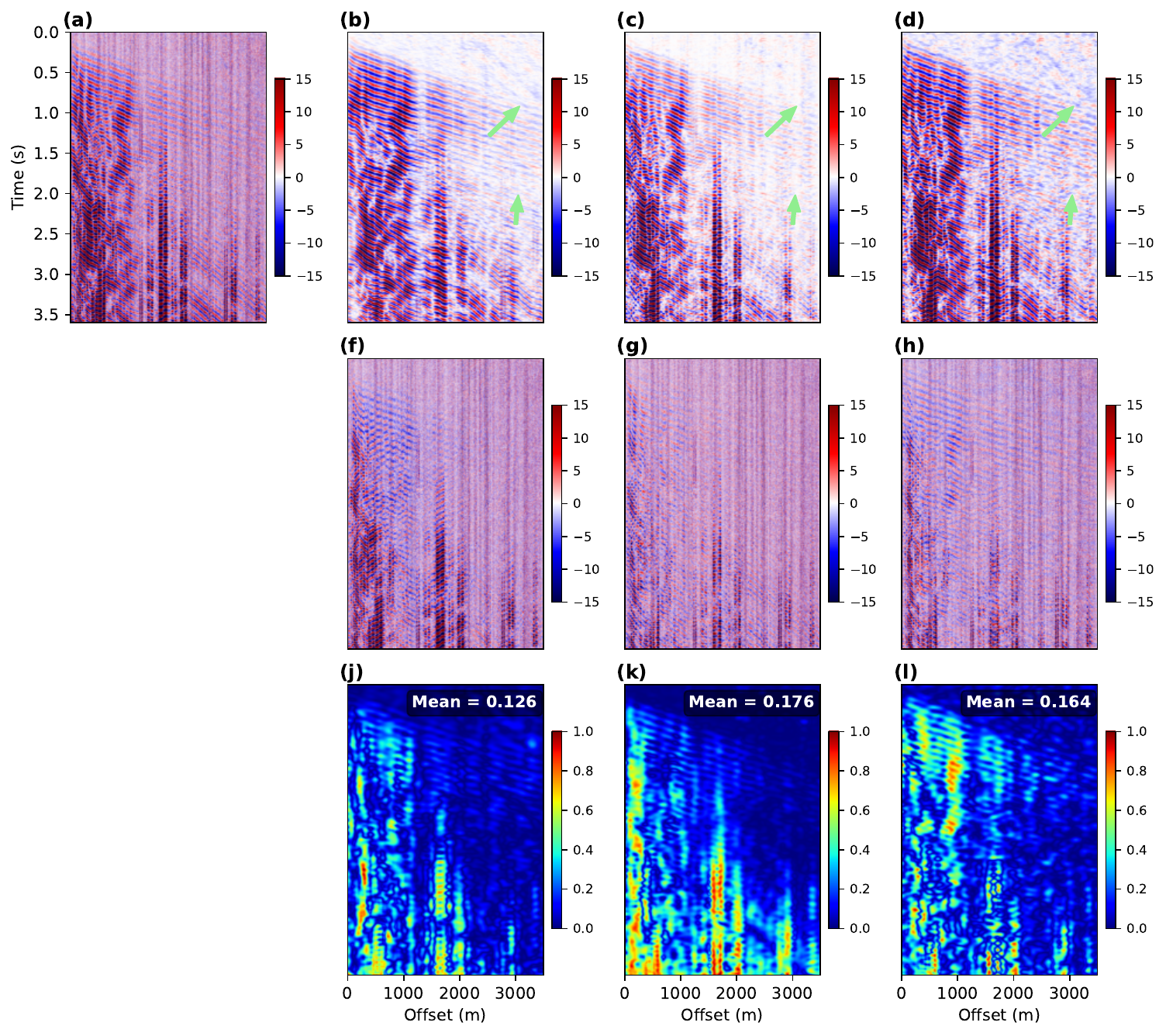}
    \caption{Out-of-distribution test on seismic data from the Gro{\ss} Sch{"o}nebeck site. (a) shows the raw input data. (b–d) present the denoised results obtained using PAUFNO, UFNO, and U-Net, respectively. (e–g) display the corresponding removed noise sections produced by each model. (h–j) illustrate the local similarity maps computed between the denoised data and the removed noise for PAUFNO, UFNO, and U-Net, respectively. Note that we provide the mean values of each similarity map, which indicates PAUFNO causes the least signal damage.}
    \label{fig:ood_test}
\end{figure}

Figure~\ref{fig:ood_test} presents the out-of-distribution test conducted on data recorded from a different survey. Compared to the Utah FORGE dataset, this dataset contains a new type of ``zigzag'' noise that was not observed in the training data, posing additional challenges for conventional deep learning models. To further evaluate denoising performance, we employ local similarity~\citep{chen2015random}, which measures the orthogonality between the denoised signal and the removed noise. Higher local similarity values indicate greater signal leakage (i.e., more signal damage), whereas lower values reflect better signal preservation. Because of the dip differences between the seismic events at the Utah FORGE site and those at the Gro{\ss} Sch{\"o}nebeck site, the data array was flipped horizontally when applying the PAUFNO method. As shown in Figures~\ref{fig:ood_test}b–\ref{fig:ood_test}d, all methods effectively suppress the strong random noise. However, the PAUFNO predictions exhibit greater spatial consistency, particularly around the regions highlighted by the green arrows in Figure~\ref{fig:ood_test}. In contrast, the UFNO and U-Net results show more signal distortion in their corresponding removed-noise sections. The local similarity maps in Figures~\ref{fig:ood_test}h–j further confirm that PAUFNO demonstrates higher robustness and better signal preservation than the benchmark methods.

\section{Discussion}
The results demonstrate a performance comparison across experiments for various methods and provide evidence that PAUFNO can effectively enhance weak seismic signals in DAS recordings, even when trained with imperfect labels. Nevertheless, a deeper understanding of PAUFNO is necessary, particularly regarding its uncertainty and inherent limitations. Therefore, in this section, we introduce Monte Carlo (MC) Dropout~\citep{gal2016dropout} within the proposed NOs-based framework to quantify model uncertainty and further investigate the limitations of PAUFNO. 

\subsection{Uncertainty quantification}
To perform UQ experiments, we conduct $T=30$ stochastic forward passes for each input patch to estimate the predictive distribution. The MC dropout probability is set to 0.2 for both models. We specifically focus our analysis on the near-wellbore traces, where the physics-aware loss is designed to correct label imperfections.

\begin{figure}
    \centering
    \includegraphics[width=1\linewidth]{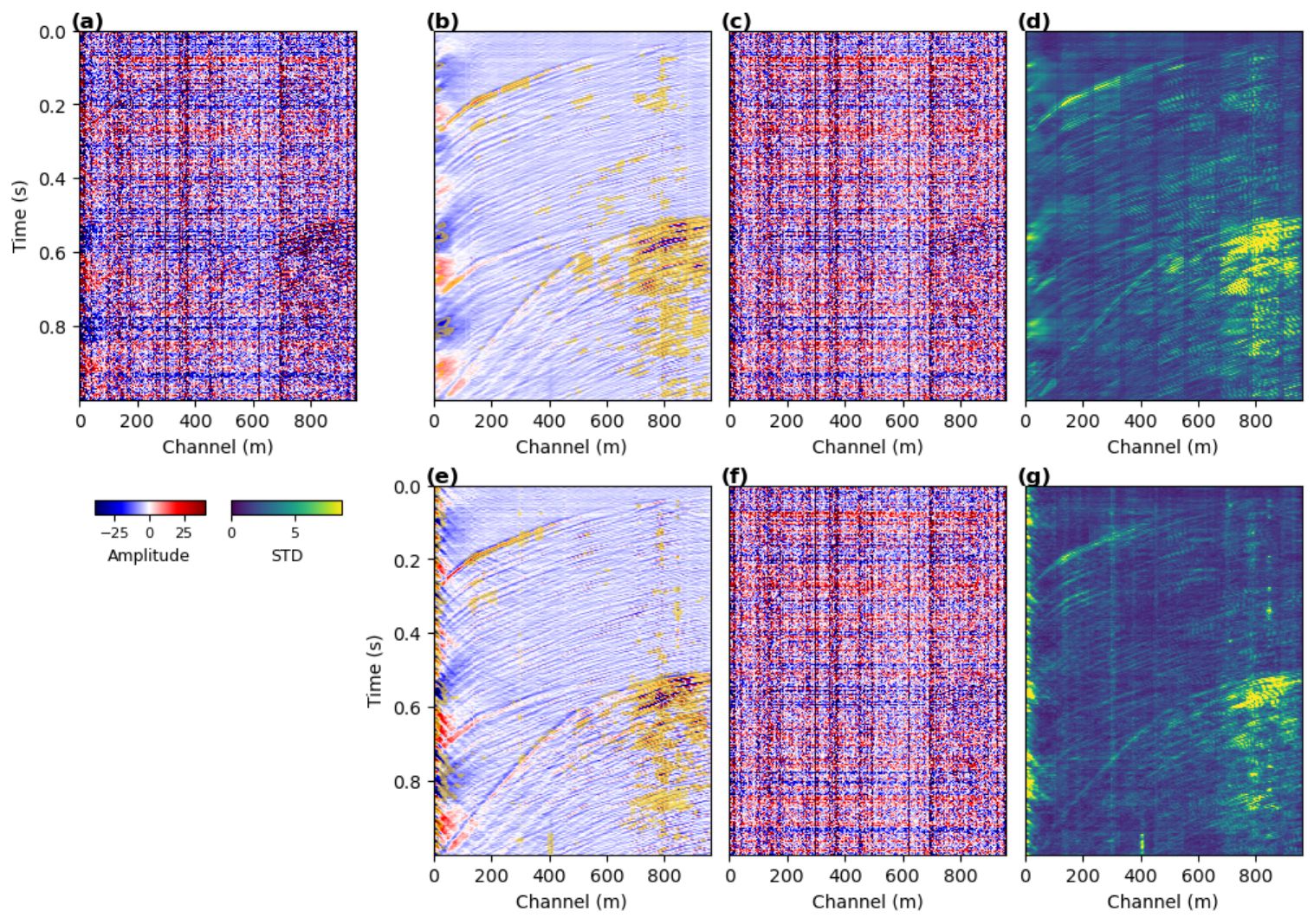}
    \caption{Side-by-side uncertainty quantification comparison of PAUFNO and UFNO on field data. Top row: (a) raw input, (b) PAUFNO prediction, (c) PAUFNO residual (raw - prediction), (d) PAUFNO predictive uncertainty (STD). Bottom row: (e) UFNO prediction, (f) UFNO residual, (g) UFNO predictive uncertainty. Seismic panels use a shared amplitude scale; uncertainty panels share the same STD scale. Contours mark the top 5\% highest uncertainty.}
    \label{fig:eq_72_uq}
\end{figure}

Figure \ref{fig:eq_72_uq} presents the MC Dropout–based UQ results for PAUFNO and UFNO. The yellow contour masks in Figures \ref{fig:eq_72_uq}b and \ref{fig:eq_72_uq}e highlight the top 5\% highest uncertainty regions. These contours indicate that PAUFNO exhibits higher confidence around the near-wellbore area, suggesting that the physics-aware loss effectively suppresses residual coupling noise in the label data. Furthermore, Figures \ref{fig:eq_72_uq}d and \ref{fig:eq_72_uq}g show the STD maps deracross experiments for various methods and provide evidence that PAUFNO can effectively enhance weak seismic signals in DAS recordings,an UFNO within channels 1–50. However, in the near-offset region, PAUFNO displays slightly higher uncertainty than UFNO, which may stem from the physics-aware loss enforcing imperfect labels too directly during training. Overall, the UQ experiments demonstrate that the proposed PAUFNO exhibits a strong robustness across the uncertainty test.

\subsection{Mask size determination}
\begin{figure}
    \centering
    \includegraphics[width=\linewidth]{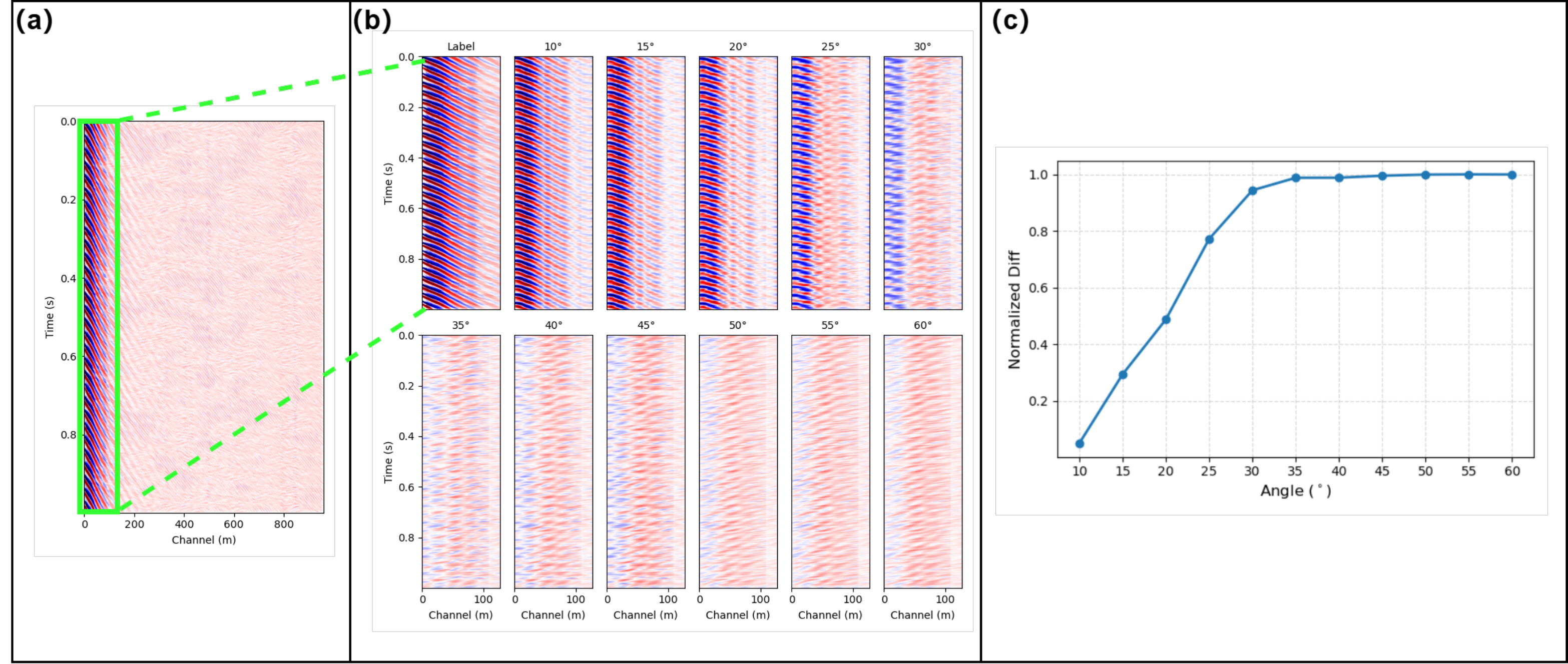}
    \caption{Illustration of selecting the short-side vertex angle~$\alpha$ (as defined in Fig.~\ref{fig:warmup_factor}a). (a) Label data from the Utah FORGE site showing strong residual coupling noise. (b) The first 128 channels of the label and predicted data obtained from models trained with dual-triangle masks using different $\alpha$ values ($10^\circ$–$60^\circ$), where the predicted sections remain dominated by residual noise. (c) Normalized mean-squared error (MSE) between the label and predictions in (b) as a function of~$\alpha$, with each value normalized by the maximum MSE for consistent comparison.}
    \label{fig:mask_selection}
\end{figure}

The physics-aware loss with the dual-triangle mask constitutes the principal novelty of this work. Consequently, investigating how to design the mask’s geometry -- specifically its size and shape is essential. We conducted extensive experiments to evaluate the influence of the short-side vertex angle, $\alpha$, on denoising performance, focusing on the first 128 channels. Figure~\ref{fig:mask_selection} illustrates the procedure for selecting the short-side vertex angle. A representative example was chosen in which no clear seismic events are visible; however, strong residual coupling noise remains, particularly within channels 0–100. Since the physics-aware loss term is designed to suppress this residual noise, analyzing the difference between the label and the model predictions provides valuable guidance for selecting an appropriate mask configuration, namely the choice of~$\alpha$.

In this analysis, the leg adjacent to $\alpha$ was fixed at a length of 40 sample points, which is sufficient to encompass the majority of the residual noise energy in the FK domain. The mean squared error (MSE) is computed as:
\begin{equation}
   MSE=\frac{1}{N} \sum_{n=1}^{N}(\hat{\mathbf{y}_{i}} - {\mathbf{y}_{i}})^2,
   \label{eq:mse}
\end{equation}
where $N = 128 \times 2000$ denotes the total number of sample points, $\hat{\mathbf{y}}$ represents the predictions obtained from models trained with different $\alpha$ values, and $\mathbf{y}$ is the corresponding label data.

Figures~\ref{fig:mask_selection}b and~\ref{fig:mask_selection}c present the predicted results and the normalized MSE curve for vertex angles ranging from $10^\circ$ to $60^\circ$. The results indicate that $\alpha = 35^\circ$ achieves satisfactory denoising performance; increasing $\alpha$ further begins to affect the signal energy adversely. Therefore, we ultimately selected the dual-triangle masks sharing a common vertex, with the vertex angle $35^\circ$, and the leg adjacent to this angle, forming the right angle, has a length of 40 sample points.

\subsection{Limitation and future work}
In the sections above, we have already demonstrated the power of the proposed UFNO and physics-aware loss function across variable experiments. However, it is necessary to point out the limitations of the proposed method to gain a deeper understanding of the proposed method, as well as provide guidance for future work. 

\begin{itemize}
    \item Patch boundary artifacts. Compared with the purely data-driven UFNO model, PAUFNO  faces challenges in handling predictions at the patch boundaries. During inference, we applied an overlap of 18 samples between adjacent patches to alleviate this issue. However, reducing the overlap leads to more noticeable “white boundary” artifacts, which can be addressed by conducting the second inference with shifted patching. Alternatively, we could only focus on the part with residual noise in which usually appears around the near-wellbore channels. 
    \item Designing triangle masks. In this work, we employed a triangular mask with a short-side vertex angle of 30° and a leg length corresponding to 40 sample points. In future research, it would be worthwhile to investigate an adaptive mask design strategy potentially guided by the spatial energy distribution or signal-to-noise ratio to further improve the model’s physical consistency and generalization capability. 
    \item To improve the imperfect labels, a physics-aware loss function is introduced to correct residual noise using appropriate masks in the FK domain. This approach allows us to focus specifically on regions containing residual noise. Once these areas are corrected, we can train another deep learning model in a purely data-driven manner, thereby mitigating the impact of imperfect labels. 
\end{itemize}

\section{Conclusions}\label{conclusion}
Recovering weak signals obscured by complex noise in DAS data remains a long-standing challenge. Conventional denoising methods often struggle to balance effective noise suppression with artifact mitigation. In this work, we proposed a new framework that incorporates a physics-aware loss function as a penalty term to correct the imperfect labels generated by filter-based approaches. In addition, a patch-based neural operator is introduced to enable high-accuracy feature representation in both the time and frequency domains. The synergy between the physics-aware loss function and operator learning establishes a novel paradigm for training with imperfect labels while achieving outstanding denoising performance. Numerical experiments demonstrate that the data-driven UFNO outperforms the U-Net, exhibiting greater flexibility in handling varying input sizes without retraining. Furthermore, by leveraging the physics-aware loss, the UFNO effectively suppresses residual coupling noise in near-wellbore channels, highlighting the robustness and accuracy of the proposed physics-informed neural operator–based denoising framework. 

This work may inspire future research in areas such as wavefield separation and other types of noise suppression, where signal and noise can be disentangled within specific transform domains (e.g., FK, Radon, or wavelet domains). In addition, incorporating the physics-aware loss term does not require the inverse of any transform, thereby mitigating artifacts and errors that typically arise during inverse transforms.

\section{Acknowledgments}
We acknowledge the support provided by King Fahd University of Petroleum and Minerals (KFUPM) through the International Summer Program Research Grant no. {ISP24201}.

\bibliographystyle{unsrtnat}
\bibliography{references}  

\end{document}